\documentclass[journal, 10pt]{IEEEtran}

\usepackage{multirow}
\usepackage[font=scriptsize,caption=false,labelsep=space]{subfig}
\usepackage{mathrsfs}
\usepackage{graphicx,float,wrapfig,epstopdf,amsmath}
\usepackage[]{algorithmicx}
\usepackage{algpseudocode,algorithm}
\usepackage{balance}
\usepackage{enumitem}
\usepackage{cuted}

\usepackage{mathtools,lipsum}
\usepackage{amsmath}
\usepackage{amssymb}
\usepackage{amsthm}
\usepackage{graphicx}
\usepackage{epstopdf}
\usepackage{times}
\usepackage{textcomp,cite}

\usepackage{url}
\usepackage{hyperref}

\usepackage{multirow}
\usepackage{threeparttable}
\graphicspath{{./figures/}}

\DeclareMathOperator*{\minimize}{minimize} 

\usepackage[table]{xcolor}
\definecolor{infocolor}{RGB}{213,229,255}
\definecolor{inteins}{RGB}{128,179,255}
\definecolor{color1}{RGB}{199,209,232}
\definecolor{color2}{RGB}{230,231,233}

\usepackage[table]{xcolor}

\usepackage{setspace}

\usepackage{enumitem}
\setlist{nolistsep}

\begin{document}

	\title{ NEAT-MUSIC: Auto-calibration of DOA Estimation for Terahertz-Band Massive MIMO Systems}

	\author{\IEEEauthorblockN{Ahmet M. Elbir, \textit{Senior Member, IEEE},  Abdulkadir Celik, \textit{Senior Member, IEEE}, \\ and Ahmed M. Eltawil, \textit{Senior Member, IEEE}  }

		\thanks{A. M. Elbir is with  the University of Luxembourg, Luxembourg, L-1855, Luxembourg; Duzce University, Duzce, 81000, Turkey; and  King Abdullah University of Science and Technology, Thuwal, 23955, Saudi Arabia (e-mail: ahmetmelbir@ieee.org).}
		
		\thanks{A. {C}elik and A. M. Eltawil are with King Abdullah University of Science and Technology, Thuwal, 23955, Saudi Arabia (e-mail:  abdulkadir.celik@kaust.edu.sa, ahmed.eltawil@kaust.edu.sa). } 
		
	}
	
	\maketitle
	\begin{abstract}
		Terahertz (THz) band is envisioned for the future sixth generation wireless systems thanks to its abundant bandwidth and very narrow beamwidth. These features are one of the key enabling factors for high resolution sensing with milli-degree level direction-of-arrival (DOA) estimation. Therefore, this paper investigates the DOA estimation problem in THz systems in the presence of two major error sources: 1) \textit{gain-phase mismatches}, which occur due to the deviations in the radio-frequency circuitry; 2) \textit{beam-squint}, which is caused because of the deviations in the generated beams at different subcarriers due to ultra-wide bandwidth. An auto-calibration approach, namely \underline{N}ois\underline{E} subsp\underline{A}ce correc\underline{T}ion technique for \underline{MU}ltiple \underline{SI}gnal \underline{C}lassification (NEAT-MUSIC), is proposed based on the correction of the noise subspace for accurate DOA estimation in the presence of gain-phase mismatches and beam-squint. To gauge the performance of the proposed approach, the Cram\'er-Rao bounds are also derived. Numerical results show  the effectiveness of the proposed approach.
	\end{abstract}
	%
	
	
	%
	
	\section{Introduction}
	\label{sec:Introduciton}
	\IEEEPARstart{T}{erahertz} (THz) band, spanning from $0.1$ to $10$ THz, has emerged as a promising frontier for the realization of significant advancements in sixth-generation (6G) wireless networks~\cite{thz_Akyildiz2022May}. Ensuring milli-degree precision in direction-of-arrival (DOA) estimation is of paramount importance to guarantee the reliability of THz sensing as well as communication applications, e.g., THz automotive radar, real-time tracking and user localization~\cite{milliDegree_doa_THz_Chen2021Aug,milliDegreeDOA_THz_Peng2016Aug,elbir2022Aug_THz_ISAC}. High-resolution DOA estimation within the THz-band, however, is impeded by myriad challenges such as high path losses, intricate propagation/scattering dynamics, and the deployment of extremely large arrays in conjunction with massive multiple-input multiple-output (mMIMO) configuration \cite{ummimoTareqOverview,thz_Akyildiz2022May,elbir2022Nov_SPM_beamforming}. To elaborate, the mMIMO systems leverage hybrid analog/digital beamforming architectures with phase shifter networks to reduce the number of radio-frequency (RF) chains.
	Nonetheless, the embedded RF circuits are susceptible to gain-phase mismatches (GPM) that necessitate periodic over-the-air estimations/calibrations, especially as these mismatches can fluctuate due to temperature variations and hardware aging~\cite{calibrationMassiveMIMOWei2020Apr}. 
	Its impact on hybrid architectures is often go overlooked and remains relatively unexamined~\cite{conver8_Hu2019Jan,calibrationMassiveMIMOWei2020Apr}.

\begin{figure}[t]
		\centering
		{\includegraphics[draft=false,width=.9\columnwidth]{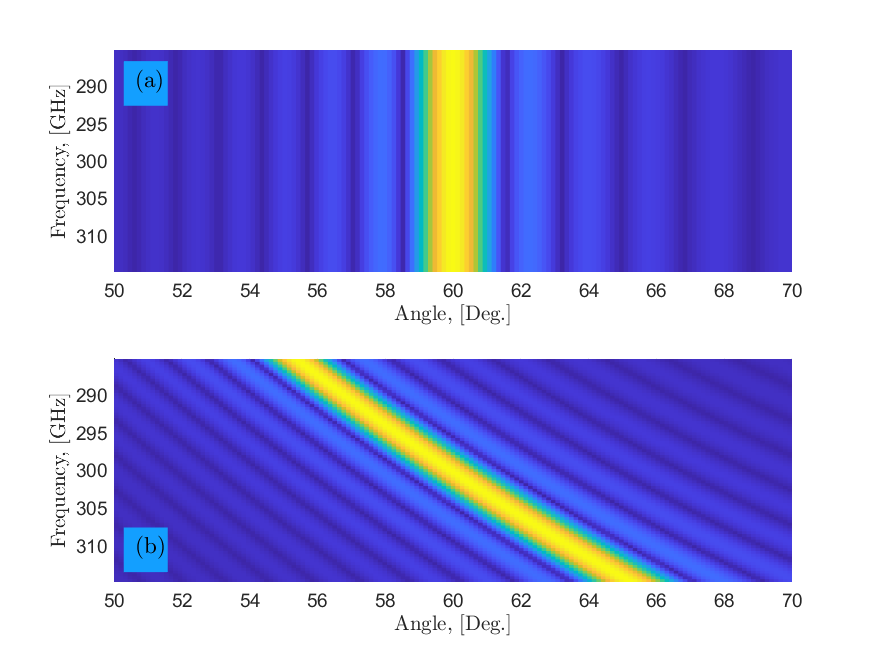} } 
		\caption{ Array gain of a single target direction at $60^\circ$ for (a) narrowband ($f_c=300$ GHz, $B = 0.1$ GHz) and (b) wideband ($f_c=300$ GHz, $B = 30$ GHz).
		}
		\label{fig_ArrayGain}
	\end{figure}

	Besides, the THz systems also suffer from \textit{beam-squint} arising from the subcarrier-independent analog beamformers~\cite{delayPhasePrecoding_THz_Dai2022Mar,beamSquint_FeiFei_Wang2019Oct,elbir_THZ_CE_ArrayPerturbation_Elbir2022Aug}. This leads to misaligned beam generation at different subcarriers squint 
	in the spatial domain; that is, the main lobes of the array gain corresponding to the lowest and highest subcarriers do not overlap because of ultra-wide bandwidth as illustrated in Fig.~\ref{fig_ArrayGain}, causing significant discrepancies in DOA estimation as a direct consequence.
	For instance, a beam-squint of roughly $6^\circ$  is observed at $0.3$ THz with a bandwidth of $30$ GHz, while it is about 
	$0.4^\circ$ for a bandwidth of $1$ GHz at $60$ GHz~\cite{elbir2022Aug_THz_ISAC,elbir_THZ_CE_ArrayPerturbation_Elbir2022Aug}. 
	Notably, existing countermeasures for beam-squint are predominantly hardware-based~\cite{thz_beamSplit}. Specifically, additional hardware components such as time-delayer networks are realized to generate a negative group-delay for its compensation~\cite{beamSquint_FeiFei_Wang2019Oct}. However, they are expensive because each phase shifter of the network is connected to multiple delayer elements, each of which consumes approximately $150\%$  more power than a single phase shifter at THz band~\cite{elbir2022Aug_THz_ISAC}. THz channel estimation~\cite{elbir_THZ_CE_ArrayPerturbation_Elbir2022Aug} and hybrid analog/digital beamforming~\cite{delayPhasePrecoding_THz_Dai2022Mar,beamSquint_FeiFei_Wang2019Oct,elbir2021JointRadarComm} under beam-squint have been explored in prior THz studies, which largely omit discussions on DOA estimation and GPM calibration.
	While the DOA estimation problem  is studied for both THz~\cite{milliDegree_doa_THz_Chen2021Aug} and millimeter-wave~\cite{calibrationMassiveMIMOWei2020Apr,doaEst_mmWave_Zhang2021Oct} mMIMO as well as phased-arrays~\cite{gpm_CL1_Dai2020Dec,gpm_WCL1_Fang2020Oct}, the impact of beam-squint is often disregarded.
	
	In this letter, we present a novel perspective, focusing on over-the-air GPM calibration of DOA estimation, especially in the context of beam-squint effects in THz mMIMO systems. 
A subspace-based auto-calibration approach is proposed, wherein the DOA angles and the GPM parameters are alternatingly estimated. Traditional subspace-based approaches, e.g.,  \textit{MU}ltiple \textit{SI}gnal \textit{C}lassification (MUSIC) algorithm~\cite{music_Schmidt1986Mar} falls short in estimating the DOAs because of inaccurate noise-subspace which is corrupted by beam-squint and GPM. {To address these issues, the main contributions of this work are as follows:
\begin{enumerate}[wide]
	\item  We introduce a \textit{N}ois\textit{E} subsp\textit{A}ce correc\textit{T}ion technique for MUSIC (NEAT-MUSIC) to estimate beam-squint-corrected DOAs. NEAT-MUSIC involves crafting a linear transformation matrix that constructs a mapping between the nominal and the beam-squint-distorted steering vectors, which facilitates the rectification of the skewed noise-subspace matrix derived from the covariance of the array data.Once the DOAs are obtained, a minimum eigenvalue problem is solved to identify the GPM parameters.
	\item  We also derive the Cram\'er-Rao bounds (CRBs) for the intended scenario to benchmark the performance of the proposed NEAT-MUSIC algorithm. 
\end{enumerate}
}


%


		
	\section{System Model}
	\label{sec:SignalModel}
	We consider a wideband THz mMIMO system, wherein the base station (BS) employs hybrid analog/digital beamformers performed over $M$ subcarriers with $N$-element uniform linear array (ULA) and $N_\mathrm{RF}$ RF chains. The BS employs  the subcarrier-independent precoder $\mathbf{F}\in \mathbb{C}^{N\times N_\mathrm{RF}}$ and the sensing signals $\mathbf{s}_m(t_i)\in \mathbb{C}^{N_\mathrm{RF}}$, where $i\in [1, T]$ and $T$ is the number of snapshots along the fast-time axis~\cite{mimoRadar_WidebandYu2019May}. To sense the environment, the BS activates $N_\mathrm{RF}$ RF chains and transmits the probing signal  $\mathbf{X}_m = \left[\mathbf{x}_m(t_1),\cdots, \mathbf{x}_m(t_T) \right]\in \mathbb{C}^{N\times T}$, where $\mathbf{x}_m(t_i) = \mathbf{F}\mathbf{s}_m(t_i)\in \mathbb{C}^{N}$, and   $\mathbb{E}\{ \mathbf{X}_m  \mathbf{X}_m^\textsf{H} \} =\frac{P_rT}{MN} \mathbf{I}_N$, for which $\mathbf{FF}^\textsf{H} = 1/N$ and $P_r$ is the transmit power.	 Assuming there are $K$ targets in the far-field of the antenna array, the received $N\times T$ target echo signal impinging on the  array is given by
	\begin{align}
	\widetilde{\mathbf{Y}}_m = \sum_{k = 1}^{K} \mathbf{G}_m \mathbf{a}(\theta_{k,m}) \widetilde{\mathbf{x}}_{k,m} + \widetilde{\mathbf{N}}_m, \label{receivedSignal1}
	\end{align}
	where  	$\widetilde{\mathbf{N}}_m \sim \mathcal{CN}(\mathbf{0},\sigma^2  \mathbf{I}_{N})
	$ is temporarily and spatially white zero-mean complex Gaussian noise matrix of size $N\times T$ with variance $\sigma^2$.  $\mathbf{G}_m = \mathrm{diag}\{\mathbf{g}_m\}$ is an $N\times N$ diagonal matrix representing the GPM parameters $\mathbf{g}_m = [g_{1,m},\cdots, g_{N,m}]^\textsf{T}\in \mathbb{C}^N$\footnote{For the mismatch-free scenario, wherein the antennas are fully-calibrated, we have $\mathbf{G}_m = \mathbf{I}_N$.}.    $\widetilde{\mathbf{x}}_{k,m}\in\mathbb{C}^{1\times T}$ denotes the echo signal reflected from the $k$-th target as $\widetilde{\mathbf{x}}_{k,m} = \beta_{m,k}\left[\mathbf{G}_m \mathbf{a}(\theta_{k,m})\right]^\textsf{T} \mathbf{X}_m$, for which $\beta_{m,k}\in \mathbb{C}$ is the reflection coefficient.  $\mathbf{a}(\theta_{k,m})\in \mathbb{C}^{N}$ is the steering vector corresponding to the physical target direction $\theta_{k} = \sin \vartheta_k$, where $\vartheta_k \in \left[-\frac{\pi}{2}, \frac{\pi}{2}\right]$, where the spatial direction $\theta_{k,m}$ is defined as $\theta_{k,m} = \eta_m \theta_{k}$. Here, $\eta_m$ denotes the distortion coefficient due to beam-squint and defined as the ratio of the subcarrier frequencies, i.e., $\eta_m = \frac{f_m}{f_c}$, where   $f_c$ is the carrier frequency and $f_m = f_c + \frac{B}{M}(m - 1 - \frac{M-1}{2})$ denotes the frequency of the $m$-th subcarrier for bandwidth $B$~\cite{delayPhasePrecoding_THz_Dai2022Mar}. To provide further insight, we define the $n$-th element of $\mathbf{a}(\theta_{k,m})$ in terms of $\theta_k$ as
	\begin{align}
	\left[\mathbf{a}(\theta_{k,m})\right]_n &= 1/\sqrt{N} \exp \big\{\mathrm{j} (n-1) \frac{2\pi d }{\lambda_m}  \theta_k   \big\} \nonumber \\
	& = 1/\sqrt{N} \exp \big\{\mathrm{j} (n-1) \pi \eta_m  \theta_k   \big\}, \label{steeringVecSquinted}
	\end{align}
	where  $d$ is the half-wavelength element spacing, i.e., $d  = \frac{c}{2f_c}$, where $c$ denotes the speed of light, and $\lambda_m = \frac{c}{f_m}$ is the wavelength of the $m$-th subcarrier\footnote{In beam-squint-free case, e.g., narrowband ($\frac{|f_m - f_c|}{f_c} \ll 1 $), we have $\eta_m \rightarrow 1$ and $\theta_{k} \rightarrow \theta_{k,m}$.}. The aim of this work is to estimate the beam-squint-corrected DOA angles $\{\theta_k\}_{k =  1}^K$ and  the GPM parameters $\{g_{n,m}\}_{n = 1}^N$ for $m\in \mathcal{M}$.


{
	\color{black}
	\begin{table}[t]
	\caption{Nomenclature
	}
	\footnotesize
	\label{tableSummary}
	\centering
	\begin{tabular} {clcl}
				\hline 		\hline 
		$\theta_k$     & True DOA &
		$\theta_{k,m}$ & Beam-squinted DOA \\
		 $\mathbf{X}_m$ & Transmitted  Signal &
		 $\mathbf{Y}_m$ & Received Signal \\
		$\mathbf{F}$   & Precoder Matrix  &
		$\mathbf{W}$  & Combiner Matrix \\
		$\mathbf{G}_m$ & GPM Matrix &
		$\mathbf{T}_m(\theta)$ & Transformation Matrix \\
		$\mathbf{U}_m^\mathrm{N}$ & Corrupted Noise Subspace  &
		$\mathbf{V}_m^\mathrm{N}$ & Corrected Noise subspace  \\
				\hline 		\hline 
		\end{tabular}
\end{table}
}


	\section{The Proposed NEAT-MUSIC Approach}
	{Define $\breve{\mathbf{W}}\in\mathbb{C}^{N\times N_\mathrm{RF}}$  as the analog combiner matrix applied to the array output $\widetilde{\mathbf{Y}}_m$ in (\ref{receivedSignal1}) as  
	\begin{align}
	\label{radarReceived1}
	\breve{\mathbf{Y}}_{m} \color{black} = \breve{\mathbf{W}}^\textsf{H}\widetilde{\mathbf{Y}}_m  = \sum_{k = 1}^{K} \breve{\mathbf{W}}^\textsf{H}\mathbf{G}_m {\mathbf{a}}(\theta_{k,m}) \widetilde{\mathbf{x}}_{k,m} + \breve{\mathbf{W}}^\textsf{H}\widetilde{\mathbf{N}}_m,
	\end{align}
	which yields an $N_\mathrm{RF}\times 1$ data for parameter estimation.
 In order to collect the full array data from $N_\mathrm{RF}$ RF chains, we follow a subarrayed approach, wherein the BS activates the antennas in a subarrayed fashion to obtain $N\times 1$ array data in $J=\frac{N}{N_\mathrm{RF}}$ time slots. Let $\mathbf{W}_j \in \mathbb{C}^{N\times N_\mathrm{RF}}$ be the applied combiner matrix at the $j$-th time slot (instead of $\breve{\mathbf{W}}$ in (\ref{radarReceived1})) as $\mathbf{W}_j = \left[
 \mathbf{0}_{jN_\mathrm{RF}\times N_\mathrm{RF}}^\textsf{T},
 \overline{\mathbf{W}}_j^\textsf{T},
 \mathbf{0}_{N-(j+1)N_\mathrm{RF}\times N_\mathrm{RF}}^\textsf{T} \right]^\textsf{T} \in \mathbb{C}^{N\times N_\mathrm{RF}}$, where $\overline{\mathbf{W}}_j\in \mathbb{C}^{N_\mathrm{RF}\times N_\mathrm{RF}}$ represents the combiner for the $j$-th block for $j = 1,\cdots, J$. Note that during collecting the received target echoes for $J = \frac{N}{N_\mathrm{RF}}$ time slots, the target DOAs are assumed to maintain invariant within a time slot while changing over time slots, which is reasonable for THz system wherein the symbol time in the order of picoseconds~\cite{milliDegree_doa_THz_Chen2021Aug,mimoRadar_WidebandYu2019May}.} 
 Then, the $N_\mathrm{RF}\times T$ echo signal reflected from the $K$ targets at the $j$-th time slot is
	\begin{align}
	\label{radarReceived}
	{\mathbf{Y}}_{j,m}  & ={\mathbf{W}_j^\textsf{H}\widetilde{\mathbf{Y}}_m}= \sum_{k = 1}^{K} \mathbf{W}_j^\textsf{H}\mathbf{G}_m {\mathbf{a}} (\theta_{k,m}) \widetilde{\mathbf{x}}_{k,m} + \mathbf{W}_j^\textsf{H}\widetilde{\mathbf{N}}_m \nonumber \\
	&	= 
	\sum_{k = 1}^K \beta_k \mathbf{W}_j^\textsf{H}{\mathbf{h}}_{k,m} {\mathbf{h}}_{k,m}^\textsf{T} {\mathbf{X}}_m + {\mathbf{N}}_{j,m},
	\end{align}
	where ${\mathbf{h}}_{k,m} = \mathbf{G}_m \mathbf{a}(\theta_{k,m})\in \mathbb{C}^{N}$ and    ${\mathbf{N}}_{j,m} = \mathbf{W}_j^\textsf{H}\widetilde{\mathbf{N}}_m\in \mathbb{C}^{N_\mathrm{RF}\times T}$ represents the noise term. Defining $\mathbf{H}_m = \left[\mathbf{h}_{1,m},\cdots, \mathbf{h}_{K,m}   \right]\in \mathbb{C}^{N\times K}$,  ${\mathbf{D}}_{j,m} = \mathbf{W}_j^\textsf{H}\mathbf{H}_m\in \mathbb{C}^{N_\mathrm{RF}\times K}$ and $\boldsymbol{\Pi}_m = \mathrm{diag}\{\beta_{1,m}, \cdots, \beta_{K,m} \}\in \mathbb{C}^{K\times K}$, (\ref{radarReceived}) becomes ${\mathbf{Y}}_{j,m} = {\mathbf{D}}_{j,m}    \boldsymbol{\Pi}_m {\mathbf{H}}_m^\textsf{T}{\mathbf{X}}_m + {\mathbf{N}}_{j,m}.$
%
	Stacking all $\mathbf{Y}_{j,m}$ into a single $N\times T$ matrix leads to the overall observation matrix $\mathbf{Y}_m\in \mathbb{C}^{N\times T}$ as $	\mathbf{Y}_m = \left[ \mathbf{Y}_{1,m}^\textsf{T}, \cdots, \mathbf{Y}_{J,m}^\textsf{T} \right]^\textsf{T} $, i.e.,
	\begin{align}
	\mathbf{Y}_m= \mathbf{D}_m \boldsymbol{\Pi}_m \mathbf{H}_m^\textsf{T} \mathbf{X}_m + {\mathbf{N}}_m, \label{obs1}
	\end{align}
	where $\mathbf{D}_m = \left[\mathbf{D}_{1,m}^\textsf{T},\cdots, \mathbf{D}_{J,m}^\textsf{T} \right]^\textsf{T} = \mathbf{W}^\textsf{H}\mathbf{H}_m \in \mathbb{C}^{N\times K}$ $\mathbf{W} = \left[{\mathbf{W}}_1,\cdots, {\mathbf{W}}_J  \right]\in \mathbb{C}^{N\times N}$ and ${\mathbf{N}}_m = \left[{\mathbf{N}}_{1,m}^\textsf{T},\cdots, {\mathbf{N}}_{J,m}^\textsf{T}  \right]^\textsf{T}$. The $N\times T$ array output data in (\ref{obs1}) is collected via limited number of RF chains from  multiple time-slots, which can be used to construct the covariance matrix to invoke the MUSIC algorithm. In the following, we introduce an alternating approach, wherein the DOA angles and GPM parameters are estimated one-by-one iteratively.
	
	\begin{algorithm}[t]
		\begin{algorithmic}[1] 
			\caption{ \bf NEAT-MUSIC}
			\Statex {\textbf{Input:}  $\mathbf{Y}_m$, $\mathbf{W}$, $K$, $\Psi$, $\overline{\epsilon}$, $\eta_m$ for $m\in \mathcal{M}$.} \label{alg}
			\State \textbf{Initialize:} $\ell=1$, $\mathbf{G}_m^{\ell} = \mathbf{I}_N$ for $m\in \mathcal{M}$. 
			\State  $\mathbf{R}_m= \frac{1}{T}{\mathbf{Y}}_m {\mathbf{Y}}_m^\textsf{H}$ for $m\in \mathcal{M}$.
			\State   Obtain the noise subspace $\mathbf{U}_m^\mathrm{N}$ from $\mathbf{R}_m$ for $m\in \mathcal{M}$. 
			\State  \textbf{for} $\theta \in \Psi$ \textbf{do}
			\State  \indent  Construct $\mathbf{T}(\theta_{m}) = \mathrm{diag}\{\boldsymbol{\tau}(\theta_{m})\}$ with  	\par  \indent $\tau_n(\theta_{m}) = \exp\{ \mathrm{j}\pi (n-1  )\Delta_m(\theta) \}$ for $m\in\mathcal{M}$.
			\State   \textbf{end for}
			\State \textbf{while} $\epsilon^\ell < \overline{\epsilon}$  \textbf{do}
			\State \indent \textbf{for} $m\in\mathcal{M}$ 
			\State
			\indent \indent Construct the corrected noise subspace as \par 
			\indent 		\indent  ${\mathbf{V}_m^\mathrm{N}}^{\ell} \gets \mathbf{T}^\textsf{H}(\theta_m)  {\mathbf{G}_m^{\ell}}^\textsf{H} \mathbf{W} {\mathbf{U}_m^\mathrm{N}}$.
			\State\indent \indent Construct the MUSIC spectra for $m$ \par 
			\State \indent \indent $P_m^{\ell}(\theta) \gets \frac{1}{\mathbf{a}^\textsf{H}(\theta){\mathbf{V}_m^\mathrm{N}}^{\ell}{{\mathbf{V}_m^\mathrm{N}}^{\ell}}^\textsf{H} \mathbf{a}(\theta)}.$
			\State \indent \textbf{end}  
			\State \indent Combined MUSIC spectra: ${P}^{\ell}(\theta)  \gets \sum_{m = 1}^M {P}_m^{\ell}(\theta)$.
			\State \indent Find $\{\hat{\theta}_k^{\ell}\}_{k =1}^{K}$ from the $K$ highest peaks of ${P}^{\ell}(\theta)$.
			\State \indent Construct $\mathbf{a}(\hat{{\theta}}_k^{\ell})$, and  solve (\ref{findGPM}) for ${\mathbf{g}}_m^\ell$ for $m\in \mathcal{M}$.
			\State \indent $\mathbf{G}_m^{\ell+1} \gets \mathrm{diag}\{ \mathbf{g}_m^\ell \}$ for $m\in\mathcal{M}$.
			\State \indent $\ell \gets \ell + 1$.
			\State \indent  $ \epsilon^\ell \gets \sum_{k = 1}^K |\hat{\theta}_k^{\ell} - \hat{\theta}_k^{\ell-1}  |$.
			\State \textbf{end while}
			
			\Statex {\textbf{Return:} $\hat{\theta}_k =\hat{{\theta}}_k^{\ell-1} $ and $\hat{\mathbf{G}}_m = \mathbf{G}_m^{\ell}$ for $m\in \mathcal{M}$.}
		\end{algorithmic} 
	\end{algorithm}
	
	\subsection{DOA Estimation}
	To estimate the physical DOA angles via NEAT-MUSIC, we first introduce the corrected noise subspace for beam-squint
	\cite{music_Schmidt1986Mar}. Define the ${N\times N}$ covariance matrix of the observations in (\ref{obs1}) as $\mathbf{R}_m= \frac{1}{T}{\mathbf{Y}}_m {\mathbf{Y}}_m^\textsf{H}$, i.e., 
	\begin{align}
	\mathbf{R}_m & = \frac{1}{T} {\mathbf{D}}_m \left( \frac{P_rT}{MN}\widetilde{\boldsymbol{\Pi} }_m\right) {\mathbf{D}}_m^\textsf{H} +  \frac{1}{T}{\mathbf{N}}_m{\mathbf{N}}_m^\textsf{H} \nonumber\\
	& \approxeq \frac{P_r}{MN} {\mathbf{D}}_m \widetilde{\boldsymbol{\Pi}}_m {\mathbf{D}}^\textsf{H}_m  +  \frac{\sigma^2}{N}  \mathbf{I}_{{N}}, \label{R_m1}
	\end{align}
	where ${\mathbf{N}}_m{\mathbf{N}}_m^\textsf{H} \approxeq \frac{{\sigma}^2 T}{N} \mathbf{I}_{N} $ since $\mathbf{W}^\textsf{H}\mathbf{W} = \frac{1}{N}$ and $	\widetilde{\boldsymbol{\Pi} }_m=  \boldsymbol{\Pi}_m\mathbf{H}_m^\textsf{T}\mathbf{H}_m^*\boldsymbol{\Pi}_m^\textsf{H} \in \mathbb{C}^{K\times K}$. The eigendecomposition of $\mathbf{R}_m$ yields $	\mathbf{R}_m = \mathbf{U}_m \boldsymbol{\Sigma}_m \mathbf{U}_m^\textsf{H},$
	where $\boldsymbol{\Sigma}_m\in \mathbb{C}^{N\times N}$ is a diagonal matrix composed of the eigenvalues of $\mathbf{R}_m$ in a descending order; $\mathbf{U}_m = \left[\mathbf{U}_{m}^\mathrm{S}\hspace{2pt} \mathbf{U}_m^\mathrm{N} \right]\in \mathbb{C}^{N\times N}$ corresponds to the eigenvector matrix; $\mathbf{U}_m^\mathrm{S}\in\mathbb{C}^{N\times K}$ and $\mathbf{U}_m^\mathrm{N}\in \mathbb{C}^{N\times \color{black} (N-K)}$ are the signal and noise subspace eigenvector matrices, respectively. 
	
	By exploiting the orthogonality of the signal and noise subspaces, i.e.,  $	\mathbf{U}_m^\mathrm{N} \perp  \mathbf{U}_m^\mathrm{S}$, and the fact that  the columns of $ \mathbf{U}_m^\mathrm{S}$ and $\mathbf{D}_m$ span the same subspace~\cite{music_Schmidt1986Mar,friedlander}, we have
	\begin{align}
	\| \mathbf{d}_{k,m}^\textsf{H}{\mathbf{U}_m^\mathrm{N}}  \|_2^2=0, \label{perp1}
	\end{align}
	where  $\mathbf{d}_{k,m} = \mathbf{W}^\textsf{H} \mathbf{G}_m\mathbf{a}(\theta_{k,m}) \in \mathbb{C}^{N} $ is the $k$-th column of $\mathbf{D}_m\in \mathbb{C}^{N\times K}$. Notice that (\ref{perp1}) implies the orthogonality with the corrupted steering vector $\mathbf{d}_{k,m}$, whereas our aim is to estimate the beam-squint-free physical DOA $\theta_k$. Therefore, we define  ${\mathbf{V}_m^\mathrm{N}}\in \mathbb{C}^{N\times \color{black} (N-K)}$ as the beam-squint-corrected noise subspace matrix, which is orthogonal to the nominal steering vectors. To that end, we first define the beam-squint transformation matrix $\mathbf{T}(\theta_{k,m})\in \mathbb{C}^{N\times N}$, which provides a linear mapping between the nominal and beam-squint-corrupted steering vectors\footnote{{Note that $\mathbf{T}(\theta_{k,m})$ only involves the corruptions due to beam-squint whereas the remaining  uncertainties can be modeled in the GPM matrix $\mathbf{G}_m$.}} as 
	\begin{align}
	\mathbf{a}(\theta_{k,m}) = 	\mathbf{T}(\theta_{k,m})\mathbf{a}(\theta_k), \label{transformation}
	\end{align}
	where $\mathbf{T}(\theta_{k,m}) = \mathrm{diag}\{\boldsymbol{\tau}(\theta_{k,m})  \}$, 	for which the $n$-th element of $\boldsymbol{\tau}(\theta_{k,m})\in \mathbb{C}^N$ is  $	{\tau_{n}(\theta_{k,m}) = \exp \{ \mathrm{j}\pi (n-1  ) \Delta_{m}(\theta_k) \} } $,
	where $\Delta_{m}(\theta_k) $ denotes the beam-squint~\cite{delayPhasePrecoding_THz_Dai2022Mar,elbir_THZ_CE_ArrayPerturbation_Elbir2022Aug} as $	\Delta_{m}(\theta_k) = (1 - \eta_m )  \theta_{k}.$
	Using (\ref{transformation}),  (\ref{perp1}) is rewritten as
	\begin{align}
	\| \left(  \mathbf{W}^\textsf{H} \mathbf{G}_m{\mathbf{a}}({\theta}_{k,m})  \right)^\textsf{H} {\mathbf{U}_m^\mathrm{N}}\|_2^2 = & \nonumber \\
	\|  \mathbf{a}^\textsf{H}({\theta}_{k}) {\mathbf{T}^\textsf{H} (\theta_{k,m}) \mathbf{G}_m^\textsf{H} \mathbf{W} {\mathbf{U}_m^\mathrm{N}}}\|_2^2 = 
	\|  \mathbf{a}^\textsf{H} ({\theta}_{k})\mathbf{V}_m^\mathrm{N}     	 \|_2^2 =& 0, \label{perp2}
	\end{align}
	where 
	$\mathbf{V}_m^\mathrm{N} \triangleq   \mathbf{T}^\textsf{H}(\theta_{k,m})  \mathbf{G}_m^\textsf{H} \mathbf{W} {\mathbf{U}_m^\mathrm{N}}$
	is the corrected noise subspace matrix. Examining (\ref{perp2}) reveals the useful property regarding the orthogonality of the corrected noise subspace $\mathbf{V}_m^\mathrm{N} $ and the  beam-squint-free steering vectors as $\mathbf{a}(\theta_k) \perp \mathbf{V}_m^\mathrm{N} $ for $m\in \mathcal{M}$. Consequently, given $\mathbf{G}_m$, we can write the beam-squint-corrected MUSIC spectra for $M$ subcarriers as 
	\begin{align}
	\label{musicSpectra2}
	P(\theta) = \sum_{m = 1}^M  \frac{1}{\mathbf{a}^\textsf{H}(\theta)\mathbf{V}_m^\mathrm{N}{\mathbf{V}_m^\mathrm{N}}^\textsf{H} \mathbf{a}(\theta)},
	\end{align}
	whose $K$ highest peaks correspond to the physical target directions $\{\hat{{\theta}}_k\}_{k=1}^K$, which can be identified through a peak-finding algorithm for (\ref{musicSpectra2}) only once since it includes the combination of spectra for $M$ subcarriers.

	\subsection{GPM Parameter Estimation}
	Using the DOA angles $\{\hat{{\theta}}_k\}_{k=1}^K$ obtained from (\ref{musicSpectra2}), the GPM parameters are found by solving the following  optimization problem, i.e., 
	\begin{align}
	\minimize_{{\mathbf{g}}_m  }  {\mathbf{g}}_m^\textsf{H} \boldsymbol{\Theta}_m{\mathbf{g}}_m , \label{findGPM}
	\end{align}
	where  $ \boldsymbol{\Theta}_m$ is an ${N\times N}$ diagonal matrix as $	\boldsymbol{\Theta}_m = \sum_{k = 1}^{K} \mathrm{diag}\{\mathbf{a}(\hat{\theta}_k)\}^\textsf{H} \mathbf{V}_m^\mathrm{N}{\mathbf{V}_m^\mathrm{N}}^\textsf{H} \mathrm{diag}\{\mathbf{a}(\hat{\theta}_k) \}.$
	The problem in (\ref{findGPM}) is convex, and its optimal solution is given by{~\cite{friedlander}
}	\begin{align}
	\hat{{\mathbf{g}}}_m = \mathrm{ev}_\mathrm{min}\{\boldsymbol{\Theta}_m \},
	\end{align}
	where $\mathrm{ev}_\mathrm{min}\{ 	\boldsymbol{\Theta}_m\}$ is the eigenvector corresponding to the smallest eigenvalue of $	\boldsymbol{\Theta}_m$.
	
	Algorithm~\ref{alg} presents the algorithmic steps for the proposed NEAT-MUSIC approach. Specifically, we first compute the beam-squint-corrupted noise subspace $\mathbf{U}_m^\mathrm{N}$ and the beam-squint transformation matrix $\mathbf{T}(\theta_{k,m})$ for $\theta\in \Psi = [-1,1]$ in Steps 2-6. Then, the estimated DOA angles $\theta_k^{\ell}$ and the GPM parameters $\mathbf{G}_m^{\ell}$ are computed iteratively in Steps  8-13 and 14-15, respectively. The alternating algorithm in NEAT-MUSIC terminates when the estimated DOA angles in two consecutive iterations satisfy $\sum_{k = 1}^K |\hat{\theta}_k^{\ell} - \hat{\theta}_k^{\ell-1}  | \geq \overline{\epsilon}$ for a pre-defined threshold $\overline{\epsilon}$.  While the alternating algorithm does not guarantee optimality, its convergence is shown in the relevant literature~\cite{friedlander,conver8_Hu2019Jan,gpm_CL1_Dai2020Dec,gpm_WCL1_Fang2020Oct}. Nevertheless, the proposed approach attains the CRB very closely (see Fig.~\ref{fig_DOA_RMSE_SNR}).

	\subsection{Computational Complexity and Identifiability}
	The complexity of the proposed NEAT-MUSIC approach is mainly due to eigendecomposition of $\mathbf{R}_m$ ($O(MN^3)$), solving the minimum eigenvalue problem in (\ref{findGPM}) ($O(MN^3)$) as well as the computation the corrected noise subspace $\mathbf{V}_m^\mathrm{N}$ ($O(MN^2[3N -K])$) for $m\in \mathcal{M}$. Thus, the overall computational complexity order is $O(MN^2[5N-K])$. Note that the complexity reduces to  $O(2MN^3)$ for the traditional MUSIC algorithm, which does not account for beam-squint. The  problem of DOA and GPM parameter estimation involves $K$ and $MN$ unknowns, respectively, while the collected array data from $N_\mathrm{RF}$ RF chains for $J = \frac{N}{N_\mathrm{RF}}$ time-slots is $N\times 1$ for $M$ subcarriers.  Hence, the proposed NEAT-MUSIC technique  is feasible only if $\mathrm{rank} \{\mathbf{U}_m^\mathrm{N}{\mathbf{U}_m^\mathrm{N}}^\textsf{H} \} = N - K \geq 1$, provided that $T \geq K$ data snapshots are available. This condition becomes $N_\mathrm{RF}-K \geq 1$ if the  output for a single time-slot is used.

	\subsection{Theoretical Performance Analysis}
	\label{sec:CRB}
	We derive the theoretical mean-squared-error (MSE) of the  DOA and GPM estimation in the presence of beam-squint. While the CRB is derived for various DOA estimation settings including narrowband~\cite{crbStoicaNehorai}, wideband~\cite{widebandDOA_CRB_Liang2019Nov} and  GPM~\cite{friedlander}, our formulation includes the beam-squint scenario. Let us first define the unknown vector as $	\boldsymbol{{\psi}} = [\theta_1,\cdots,\theta_K,\Delta_{1}(\theta_1),\cdots,\Delta_{M}(\theta_K),\mathbf{g}_1^\textsf{T},\cdots, \mathbf{g}_M^\textsf{T}]^\textsf{T}\in \mathbb{C}^{Q}$, where $Q = K + MK+MN$. Then, the MSE for $\psi_i$ is lower bounded as $	\mathbb{E}\{(\hat{\psi}_i - \psi_i)(\hat{\psi}_i - \psi_i)^* \} \geq [\mathbf{CRB}]_{ii},$
	where $i \in [1, Q]$ and  $\mathbf{CRB}\in \mathbb{C}^{Q\times Q}$ denotes the the CRB matrix whose inverse has the following relationship with the Fisher information matrix as $		[\mathbf{CRB}^{-1}]_{ij} = [\mathbf{FIM}]_{ij},$
	where $i,j \in [1, Q]$.	To obtain $\mathbf{FIM}$, we compute the logarithm of the joint probability density function for $T$ statistically independent observations of $\mathbf{Y}_m =[ \mathbf{y}_m(t_1),\cdots, \mathbf{y}_m(t_T)]$ as $	\mathcal{L} = \ln \{p(\mathbf{y}_m(t_1)),\cdots, p(\mathbf{y}_m(t_T)) \} 
	 $$= - T \ln \left\{ |\boldsymbol{R}_m|\right\} - T \mathrm{Tr}\{\boldsymbol{R}_m^{-1}\mathbf{R}_m \},$
	where $\boldsymbol{R}_m = \mathbb{E}\{\mathbf{Y}_m\mathbf{Y}_m^\textsf{H}\}$ is the true covariance matrix and $\mathrm{Tr}\{\cdot\}$ is the trace operation. Then, the $\mathbf{FIM}$ is be computed from the expected value of the second derivative of $\mathcal{L}$ with respect to (w.r.t.) $\boldsymbol{{\psi}}$~\cite{crbStoicaNehorai,friedlander} as
	\begin{align}
	[\mathbf{FIM}]_{ij}& \hspace{-3pt}= - \mathbb{E} \left\{\frac{\mathcal{L}}{\partial\psi_i \partial \psi_j } \right\} \hspace{-3pt}= T \mathrm{Tr}\left\{ \boldsymbol{R}_m^{-1} \frac{\partial \boldsymbol{R}_m}{\partial \psi_i}\boldsymbol{R}_m^{-1} \frac{\partial \boldsymbol{R}_m}{\partial \psi_j}  \right\}, \nonumber 
	\end{align}
	where $\mathbb{E}\{\mathbf{R}_m  \} = \boldsymbol{R}_m$.
	Following the steps in~\cite{crbStoicaNehorai,widebandDOA_CRB_Liang2019Nov}, we get $	[\mathbf{CRB}]_{ij} =  \frac{\sigma^2}{2T}\sum_{m = 1}^{M} \frac{1}{\mathrm{Tr}\{\mathbf{K}_m\boldsymbol{\Xi}_m^{ij}  \}   },$
	where $		\mathbf{K}_m = {\widetilde{\boldsymbol{\Pi}}}_m^\textsf{H}\mathbf{D}_m^\textsf{H} \boldsymbol{R}_m^{-1} \mathbf{D}_m{\widetilde{\boldsymbol{\Pi}}}_m \in \mathbb{C}^{K\times K}$ and  $\boldsymbol{\Xi}_m^{ij}\in\mathbb{C}^{K\times K}$ include the derivative of the actual steering matrix $\mathbf{D}_m = \mathbf{W}^\textsf{H}\mathbf{H}_m = \mathbf{W}^\textsf{H} \left[\mathbf{h}_{1,m},\cdots, \mathbf{h}_{K,m} \right]$ w.r.t. $\boldsymbol{{\psi}}$ as $	\boldsymbol{\Xi}_m^{ij} = \left\{\frac{\partial \mathbf{H}_m }{\partial \psi_i} \right\}^\textsf{H} \mathbf{W} \left(\mathbf{I}_N - \mathbf{D}_m\mathbf{D}_m^\dagger   \right)\mathbf{W}^\textsf{H} \left\{\frac{\partial \mathbf{H}_m }{\partial \psi_j} \right\},$
	where $i,j \in [1, N]$ and $(\cdot)^\dagger$ denotes the More-Penrose pseudo inverse.
	To construct $\boldsymbol{\Xi}_m^{ij} $, we need to compute the derivatives of $n$-th element of $\mathbf{h}_{k,m} = \mathbf{G}_m \mathbf{a}(\theta_{k,m})$, i.e., $[\mathbf{h}_{k,m}]_n =  g_{n,m} e^{\mathrm{j}\pi ({n}-1) \eta_m\sin\tilde{\theta}_k} $, w.r.t. $\theta_k$, $\Delta_m(\theta_k)$ and $g_{n,m}$ for $n \in [1, N]$, respectively, as
	\begin{align}
	\frac{\partial [\mathbf{h}_{k,m}]_{n}}{\partial \theta_k } &= \mathrm{j} g_{{n},m}   {   \pi ({n} - 1 ) \eta_m} \cos\tilde{\theta}_k [\mathbf{h}_{k,m}]_{n}, \\
	\frac{\partial [\mathbf{h}_{k,m}]_{n}}{\partial \Delta_m(\theta_k)}\hspace{-2pt} &= \mathrm{j}g_{{n},m}     { \pi ({n}\hspace{-2pt} - \hspace{-2pt}1 )\hspace{-2pt}\frac{\eta_m}{1-\eta_m} } e^{\mathrm{j}\pi ({n}-1) \eta_m{\theta}_k}\hspace{-2pt} [\mathbf{h}_{k,m}]_{n},  \\
	\frac{\partial [\mathbf{h}_{k,m}]_{n}}{\partial g_{n,m}   }	&=   [\mathbf{a}(\theta_{k,m})]_{n}.
	\end{align}

	%

	\section{Numerical Experiments}
	{The efficiency of our NEAT-MUSIC algorithm is benchmarked against the direct application of the MUSIC algorithm~\cite{music_Schmidt1986Mar}, the MUSIC algorithm with known GPM parameters (i.e., $\mathbf{G}_m$) or known beam-squint (i.e., $\Delta_{m}(\theta_{k})$) as well as the asymptotic performance bound, i.e., CRB, which is  derived} in Sec.~\ref{sec:CRB} in terms of  root-MSE (RMSE), i.e., $\mathrm{RMSE}_\theta = (\frac{1}{J_TK} \sum_{i=1}^{J_T}\sum_{k=1}^K | \hat{{\theta}}_{i,k}- {{\theta}}_{i,k}|^2 )^{1/2}$, where $\hat{{\theta}}_{i,k}$ stands for the estimated DOA for the $i$-th instance of $J_T= 500$ Monte Carlo trials. The default simulation parameters are $f_c=300$ GHz, $B=30$ GHz, $M=32$, $N = 128$, $N_\mathrm{RF}=8$, $T=500$, $K=2$~\cite{milliDegree_doa_THz_Chen2021Aug,delayPhasePrecoding_THz_Dai2022Mar}. The DOAs are selected uniform at random from the interval $\tilde{\theta}_{k}\sim \mathrm{unif} [-\frac{\pi}{2},\frac{\pi}{2}]$. {In order to achieve asymptotic DOA estimation performance, the angular sector is divided into $2^{14}$ uniform grid points for the calculation of $\mathbf{a}(\theta)$, $\mathbf{a}(\theta_m)$ as well as $\mathbf{T}_m(\theta)$ in (\ref{musicSpectra2}).}  The GPM parameters are generated based on $\mathbf{G}_m \sim \mathcal{CN}(\mathbf{I}_N,\frac{1}{\sigma_G^2})$, with the signal-to-noise ratio (SNR) $\mathrm{SNR}_{G} = 10\log_{10}\left(\frac{1}{\sigma_G^2}\right)$. The combiner matrix is modeled as $[\mathbf{W}]_{i,j} = \frac{1}{\sqrt{N}}e^{\mathrm{j}{\phi}}$, where ${\phi} \sim \text{unif}[-\frac{\pi}{2},\frac{\pi}{2}]$ for $i \in [1, N]$ and $j  \in [1,  N_\mathrm{RF}]$. Our NEAT-MUSIC method presented in Algorithm~\ref{alg} approximately converges within $\ell = 20$ iterations for $\overline{\epsilon} = 10^{-4} $. 

	\begin{figure}[h]
		\centering
		{\includegraphics[draft=false,width=\columnwidth]{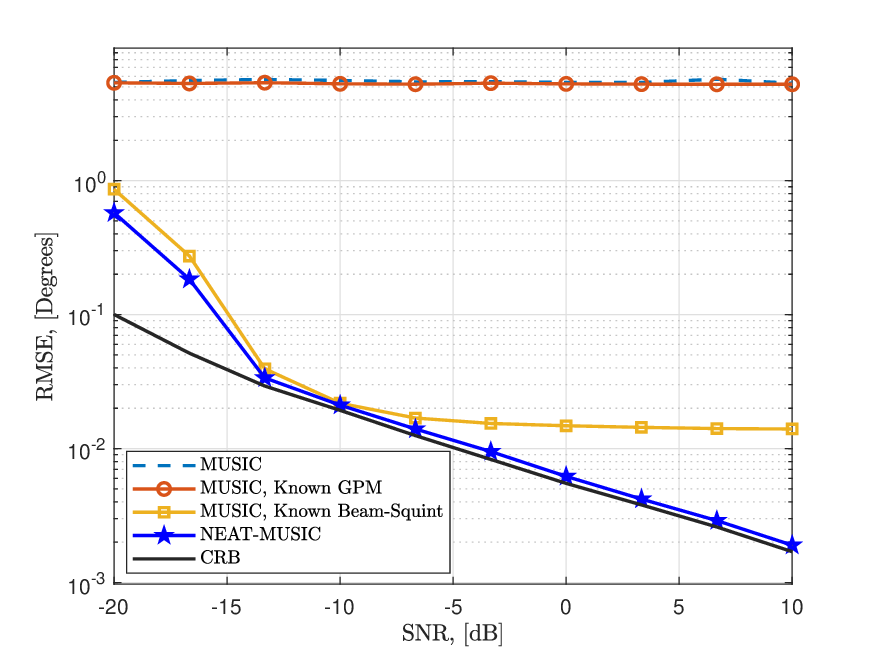} } 
		\caption{ DOA estimation RMSE vs. SNR when $\mathrm{SNR}_G = 10$ dB. 
		}
		\label{fig_DOA_RMSE_SNR}
	\end{figure}

	Fig.~\ref{fig_DOA_RMSE_SNR} shows the DOA estimation RMSE with respect to SNR, defined as $\mathrm{SNR} = 10\log_{10}(\frac{\rho}{\sigma_n^2})$ with $\rho = \frac{P_{r}}{M^2N^2} = 1$,  when the GPM parameters are generated with $\mathrm{SNR}_\mathrm{G} = 10$ dB. Notably, even the informed MUSIC algorithm displays a relatively poor performance, marked by an approximate $5^\circ$ DOA error, primarily because beam-squint effects are overlooked. 
	When beam-squint is perfectly calibrated (utilizing NEAT-MUSIC but excluding GPM calibration),  the DOA error is lower than that of known GPM case. Nevertheless, the RMSE cannot be further improved because of the precision loss due to uncalibrated GPM and yields approximately $0.02^\circ$ for $\mathrm{SNR} \geq -10$ dB. Thus, we can conclude that beam-squint causes much severer RMSE than that of GPM for DOA estimation problem. In contrast, our novel NEAT-MUSIC algorithm  outperforms the competing algorithms by attaining the CRB very closely.  This superior performance of NEAT-MUSIC can be attributed to the calibration of both GPM and beam-squint without any priori knowledge, thereby proving high resolution DOA accuracy.

	\begin{figure}[h]
		\centering
		{\includegraphics[draft=false,width=\columnwidth]{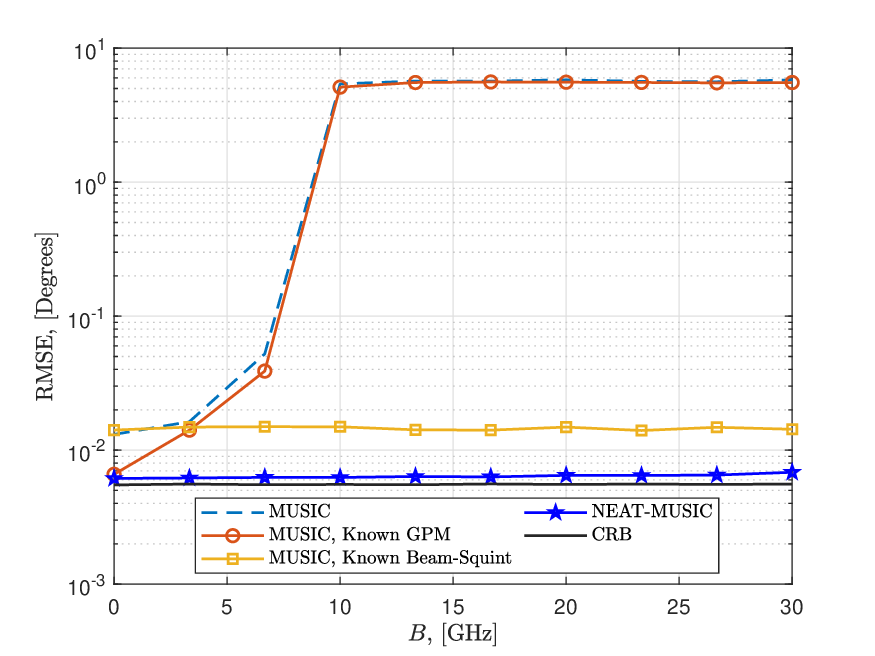} } 
		\caption{ DOA estimation RMSE vs. $B$ for  $\mathrm{SNR} = 0$ dB, $\mathrm{SNR}_G = 10$ dB.
			\vspace*{-3.5cm}}
		\label{fig_DOA_RMSE_B}
	\end{figure}
	Fig.~\ref{fig_DOA_RMSE_B}  shows the DOA estimation RMSE versus bandwidth $B$. In comparison, GPM-only and beam-squint-only calibration lead to approximately $5^\circ$ and $0.02^\circ$ DOA error while our NEAT-MUSIC algorithm provides accurate DOA estimation RMSE for wide range of bandwidth, i.e., $B\in [0,30]$ GHz.

	\section{Summary}
	In this work, we investigated the DOA estimation problem for wideband THz mMIMO system. An auto-calibration approach, called NEAT-MUSIC, is proposed to accurately estimate the DOA angles in the presence of beam-squint and GPM. While the latter has a marginal impact   ($\sim 0.02^\circ$) on DOA estimation, the former causes significant errors in the array gain and be severe ($\sim 5^\circ$).  It is shown that our NEAT-MUSIC approach can effectively estimate the DOA angles with high precision, without requiring additional hardware components, e.g., time-delayer networks.

	\bibliographystyle{IEEEtran}
	\bibliography{references_127}

\end{document}